# Enhanced piezoelectric response at nanoscale vortex structures in ferroelectrics

Xiaowen Shi[1]†, Nimish Prashant Nazirkar[1]†, Ravi Kashikar[2]†, Dmitry Karpov[3], Shola Folarin[2], Zachary Barringer[1], Skye Williams[1], Boris Kiefer[5], Ross Harder[6], Wonsuk Cha[6], Ruihao Yuan[7], Zhen Liu[8], Dezhen Xue[7], Turab Lookman[4*], Inna Ponomareva[2*], Edwin Fohtung[1*].

1 Department of Materials Science and Engineering, Rensselaer Polytechnic Institute, 110 8th St, Troy, NY 12180, USA;
2 Department of Physics, University of South Florida, 4202 East Fowler Avenue, ISA 5103, Tampa, FL 33620-5700
3 ESRF—The European Synchrotron, ID16A Beamline, 38043, Grenoble Cedex 9, France;
4 AiMaterials Research LLC, Santa Fe, New Mexico 87501, USA
5 Department of Physics, New Mexico State University, 1255 N Horseshoe, Las Cruces, NM 88003, USA
6 Advanced Photon Source, Argonne, IL 60439, USA;
7 State Key Laboratory for Mechanical Behavior of Materials, Xi'an Jiaotong University, Xi'an 710049, China
8 Department of Materials Science, Technical University of Darmstadt, Darmstadt 64287, Germany.
* Correspondence: fohtue@rpi.edu, iponomar@usf.edu, turablookman@gmail.com
†These authors contributed equally to this work.

The piezoelectric response is a measure of the sensitivity of a material's polarization to stress or its strain to an applied field. Using *in-operando* x-ray Bragg coherent diffraction imaging, we observe that topological vortices are the source of a five-fold enhancement of the piezoelectric response near the vortex core. The vortices form where several low symmetry ferroelectric phases and phase boundaries coalesce. Unlike bulk ferroelectric solid solutions in which a large piezoelectric response is associated with coexisting phases in the proximity of the triple point, the largest responses for pure $BaTiO_3$ at the nanoscale are in spatial regions of extremely small spontaneous polarization at vortex cores. The response decays inversely with polarization away from the vortex, analogous to the behavior in bulk ceramics as the cation compositions are varied away from the triple point. We use first-principles-based molecular dynamics to augment our observations, and our results suggest that nanoscale piezoelectric materials with large piezoelectric response can be designed within a parameter space governed by vortex cores. Our findings have implications for the development of next-generation nanoscale piezoelectric materials.

Piezoelectric materials have been widely utilized in a plethora of applications, including actuators, relays, and sensors [1]. More recently, piezoelectric nanocrystals have emerged as a promising platform for advanced computing, data storage, and processing, due to the ability to manipulate ferroelectric vortices and spontaneous polarization textures. The enhancement of the piezoelectric response in these nanocrystals through controlled phase transformation [2] has the potential to miniaturize piezoelectric applications, such as piezoelectric transducers and memory nano-bits[3–5]. The piezoelectric response ($d_{ij}$) was first observed by the Curie brothers in 1880[6] and the theory of piezoelectricity was later developed by W. Voigt in 1890[7] where the relationship between an applied electric field ($E_i$) and strain ($\varepsilon_i$),

$$\varepsilon_i = d_{ij}E_i \qquad \text{Eq.(1)}$$

was established. Obtaining a large $d_{33}$ desired in applications requires exploring and navigating the competition between chemical modifications (via dopants) and microstructural degrees of freedom (grain and domain size, texturing) that lead to a complex phase space.

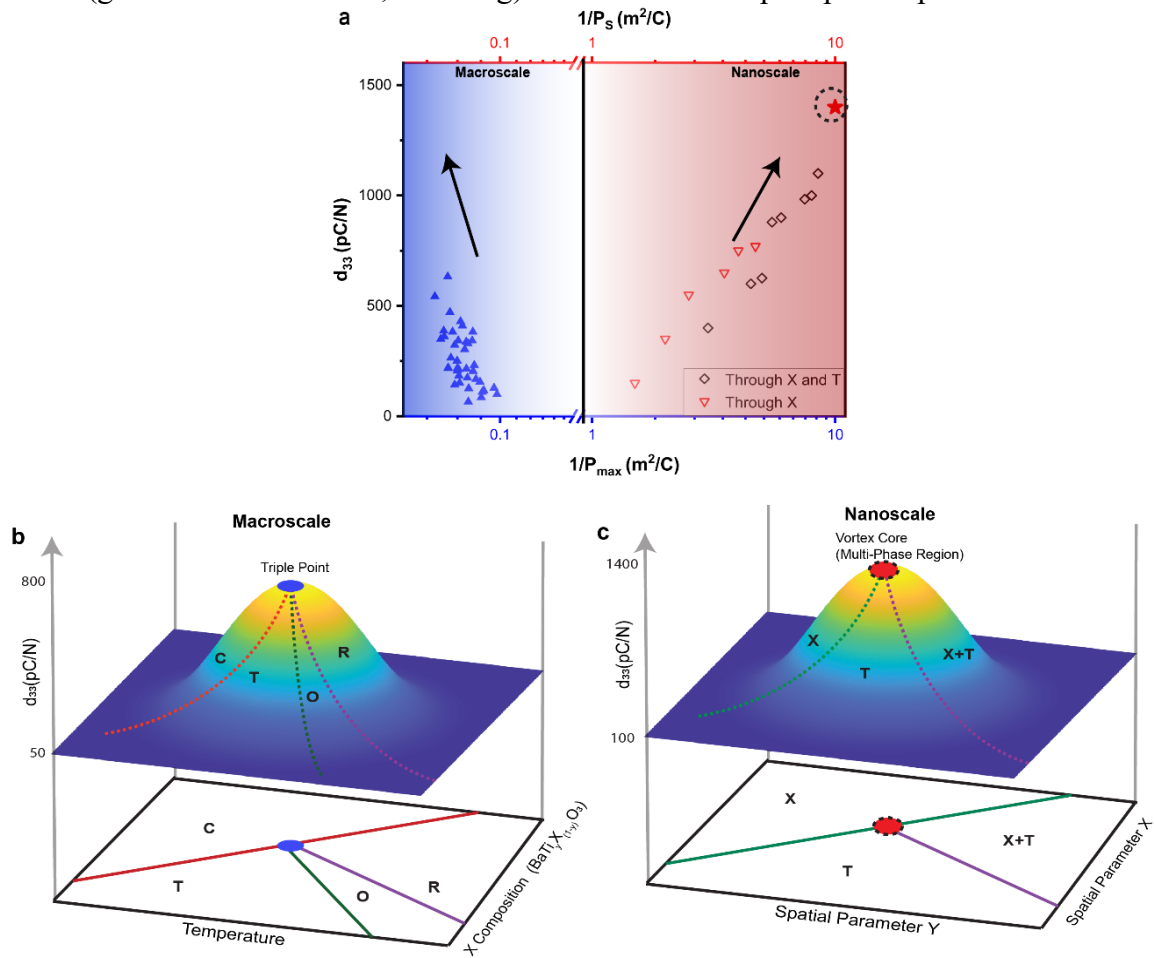

Figure 1: **Enhanced piezoelectric response and proposed structural analogue of the triple point. A:** Scatter plot showing dependence of the piezoelectric response $d_{33}$ on $1/P_{max}$, where $P_{max}$ is the maximum polarization $P_{max}$ in bulk BTO-based solid solutions reported in the literature[12,31]. The red data points show correlations between spatially measured $d_{33}$ values and the spontaneous polarization within an individual BTO nanocrystal as obtained in this work. **b:** Landscape of macroscopic piezoelectric response ($d_{33}$) distribution on the compositional phase diagram of bulk BTO-based solid solutions Ba(Ti$_y$X$_{1-y}$)O$_3$ showing that the maximum in

$d_{33}$ has been observed to occur in the multi-phase coexistence region around the triple point. **c**: Proposed nanoscale structural analogue of the phase diagram in which the vortex plays a role similar to the triple point where multiple phases coalesce.

However, it has been recognized that piezoelectrics such as Pb(Zr,Ti)O$_3$ and Pb free solutions based on BaTiO$_3$ (BTO) and BiFeO$_3$, exhibit a large piezoelectric performance in the proximity of morphotropic phase boundaries (MPB)[8–11]. Relatively large $d_{33}$ 's (~ 700 pC/N) have been observed in (Ba, Ca)(Zr, Ti)O$_3$, (Ba, Ca)(Sn, Ti)O$_3$, and (Ba, Ca)(Hf, Ti)O$_3$ using compositions (Fig. 1a) in the vicinity of the triple-point in their phase diagrams as shown in Fig. 1b. This region has been recognized as especially significant as there is a coexistence of multiple phases where MPBs terminate, and where there are potentially tri-critical points where the nature of the transition itself can change abruptly from first to second order. This region, which is "*soft*" in the presence of a field with degenerate or very low energy barriers for polarization rotation between phases, is precisely what is exploited in bulk systems to obtain a large piezoelectric response. Close to multiphase convergence in BTS$_{0.11}$-xBCT, Zhao et al.[2] observe a wide range of compositions (0.05 < x <0.22) and temperatures (10° - 40° C) where $d_{33}$ ranges between 600-700 pC/N.

Their atomic resolution polarization maps via Z contrast imaging show the rotations between the lower symmetry O, T and R phases. Kim et al[12] using TEM with a charge meter obtain an exceptionally high $d_{33}$ under compression > 1500 pC/N in BTO nanoparticles in the size range 50-120 nm. The $d_{33}$ value decreases rapidly to 260 pC/N for sizes 300 nm and 1000 nm as the bulk scale is approached. Figure 1a is a scatterplot that shows how the experimentally measured $d_{33}$ values for BTO based bulk samples, which we have synthesized and characterized in our laboratory as well as compiled from the literature, vary with the maximum polarization ($P_{max}$) seen in hysteresis data (Supplementary Section 1). The essentially linear trend between $d_{33}$ and $1/P_{max}$ in Fig.1a as dopant compositions are varied in BTO-based bulk materials is not surprising as it follows from equilibrium thermodynamics assuming that the strain, $\varepsilon$, varies with polarization as $\varepsilon \sim P^2$, a result of minimizing a generic piezoelectric energy contribution to lowest order of the form $A\varepsilon^2 + Q\varepsilon P^2$, where A and Q are related to the elastic and electrostritive constants.

Experiments[3,14–16], as well as many simulations[17–22], have beautifully established the presence of multiple phases and polarization rotations in BTO-based compositions at the nanoscale. Our previous Bragg Coherent Diffractive Imaging (BCDI) work, coupled with phase field simulations, has also picked out the monoclinic phase and shown how the phase fractions change as the vortex is driven under the influence of an external field[3]. However, lacking has been a demonstration of the specific structural features at this scale that give rise to a large $d_{33}$. This is essential if we are to learn how to engineer piezoelectric based devices with large response at the nanoscale. Moreover, if bulk samples show large $d_{33}$ close to the triple point region, can we identify the analog of the triple point at the nanoscale in ferroelectric structures where $d_{33}$ has large values which then decrease as we move away from it? We also investigate how $d_{33}$ correlates with $1/P$ at the nanoscale, and if it does, how do we interpret this equilibrium relationship we know is generally valid for bulk samples? We will explore these questions *in-operando* with Bragg coherent diffractive imaging (BCDI), a departure from the surface microscopies referred to earlier[3] as we interrogate the lattice strain response to an applied electric field within the volume of an individual BTO nanocrystal, this allows us to spatially resolve in three-dimensions the local displacements, strains, and piezoelectric $d_{33}$ response in individual 200 nm wide nanocrystals (See Fig S2 (a, b), Fig S10-12). We also utilize atomistic molecular dynamics (MD) simulations based on a first principle effective Hamiltonian[21] to model an individual BTO nanodot with lateral size 12.7 nm.

The focus of this letter is to understand the origin of the large $d_{33}$ we observe at the nanoscale and relate to the analog behavior observed at the bulk scale. We will show that the cores of the topological vortex structures where several phases and phase boundaries coalesce, is where $d_{33}$ is large but the polarization is very small. In addition, $d_{33}$ is correlated with $1/P$ but now as a function of distance away from the vortex core towards a homogeneous phase within the nanostructure. This contrasts with how $d_{33}$ varies with $1/P$ as a function of dopant compositions in bulk. Thus, we observe an analog of the bulk behavior at the nanoscale, and our findings provide a means of optimizing and engineering nanostructures with large piezoelectric response containing vortices.

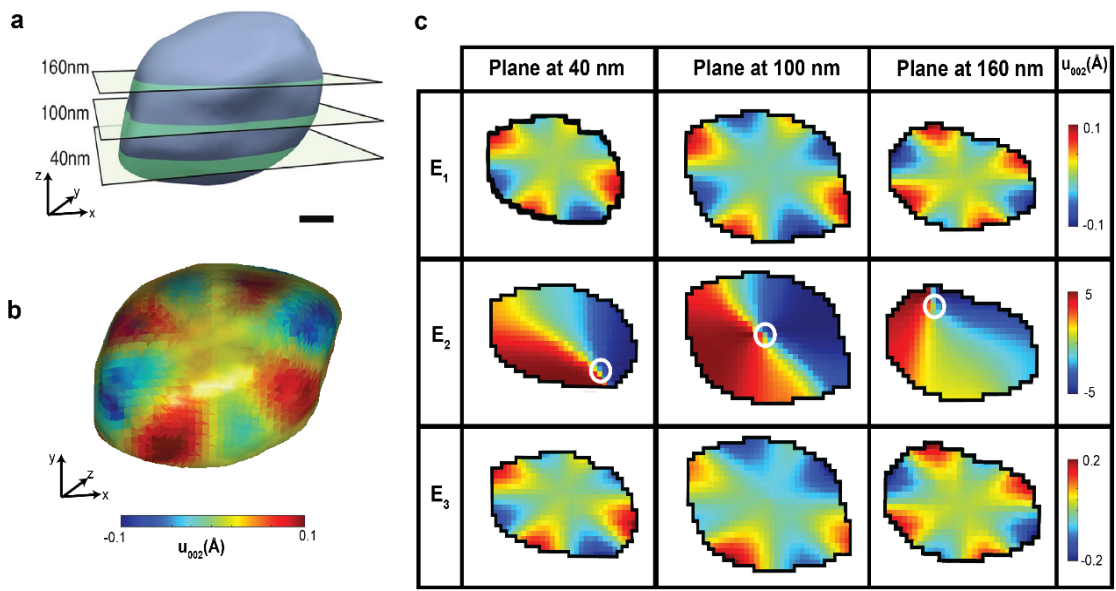

**Figure 2: Operando experimental 3D visualization of lattice distortions and observed topological vortices. a,b:** Reconstructed 3D BTO nanocrystal and a depiction of the lattice displacement field of (002) planes. **c:** scalar-cut-planes of quantitative magnitudes of displacements for electric fields $E_1$ (zero electric field at the beginning of field application), $E_2$, and $E_3$ (zero field at the end of the field removal) at several cross sections of the nanocrystal shown in panel a. The white circles depict the location of the vortex core. The scale in all figures is 20 nm.

Using BCDI, we reconstructed the shape of an individual BTO nanocrystal and resolved the atomic displacement of (002) planes for any given 3D view of the nanocrystal (Fig. 2a and b). By tracking 2D slices within the nanocrystal volume, we identified displacement patterns associated with the presence of topological ferroelectric (FE) vortices, which are also predicted by MD simulations (Fig. 4c). Fig. 2c also shows how FE vortices can be manipulated using external electric fields [3], inducing structural phase transformations within the nanocrystal (Fig. 2c). Our high volumetric spatial resolution (~16nm) of BCDI allowed us to reconstruct the evolution of the displacement field under three external fields, $E_1 = 0$ kV/cm, $E_2 = 286$ kV/cm and at remnant $E_3 = 0$ kV/cm, providing valuable insights into the behavior of FE vortices in BTO nanocrystals.

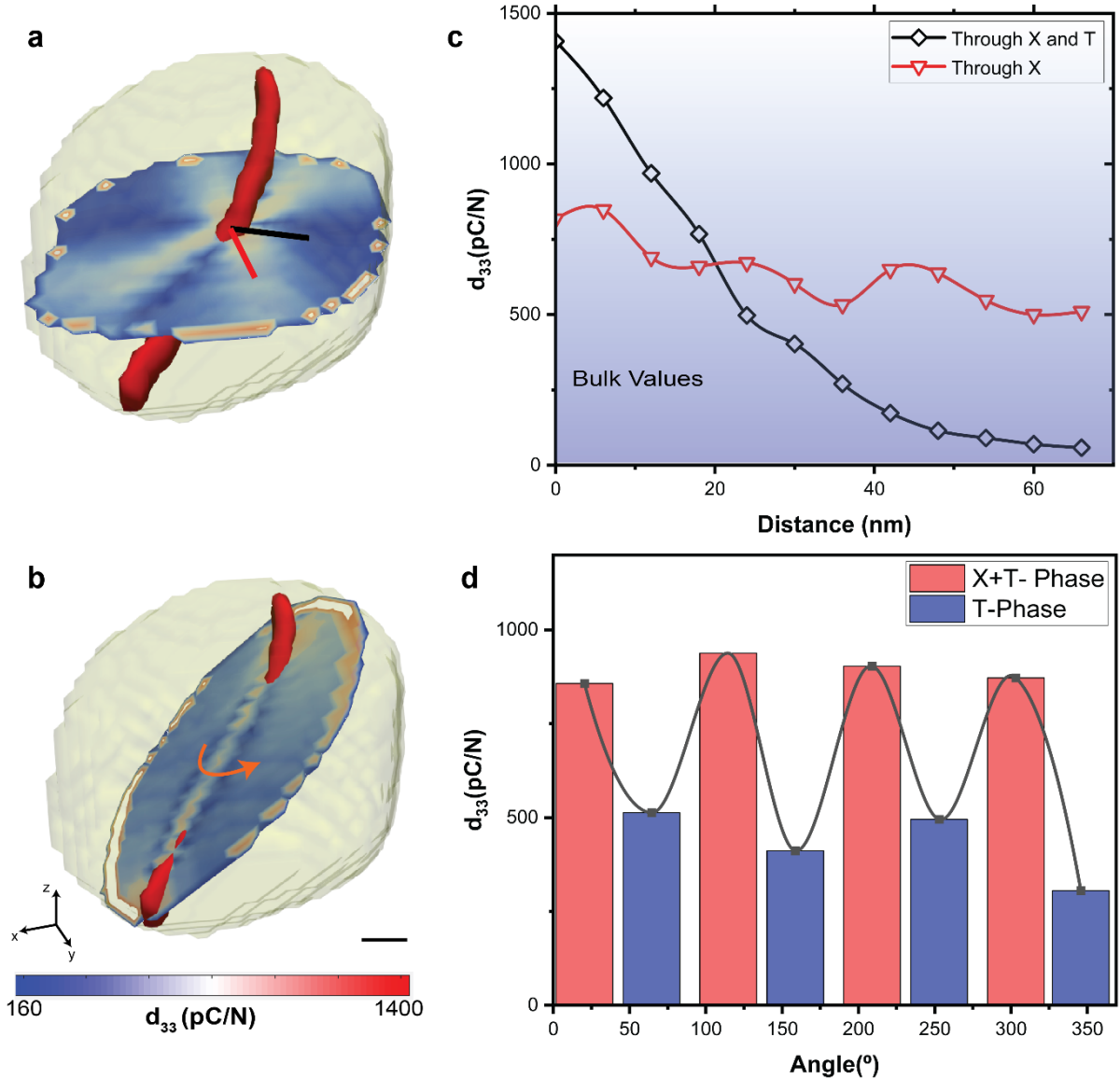

**Figure 3: Structure of topological vortex core and the enhancement of piezoelectric coefficient $d_{33}$ in BTO nanocrystal. a**: Transparent iso-surface of nanocrystal depicting a scalar cut plane of the central slice of 3D $d_{33}$ map. Structure of the vortex core is rendered as a 1D filament (red) spanning the BTO nanocrystal. The red line in **a** show a walk through the low symmetry X away from vortex core and the black line shows a walk through both X and T phases away from the vortex core.; **b:** Transparent iso-surface of nanocrystal depicting a scalar cut plane along the vortex core of 3D $d_{33}$ map. The orange arrow indicates the walk around the vortex with a radius of about 30 nm. The red region shows the iso-surface of vortex core in **a** and **b**. **c:** Linear variations of $d_{33}$ away from the vortex core into the X and T-structural phase regions (Black line in **a**) and variations along the X regions (Red line in **a**). **d:** Radial variation of the $d_{33}$ with vortex core as the center along the orange region in **b**. We observe a wave like behavior showing enhancement in the X+T-structural region(red) and a near bulk behavior in the T-structural region(blue). The scale bar represents 20 nm.

As BCDI allows us to map the displacement field along the c-direction and extract the strain tensor $\varepsilon_{33}$ as gradients $\varepsilon_{33} = \frac{du_{002}}{dz}$, this provides the ability to spatially resolve the piezoelectric coefficient $d_{33} = \frac{\varepsilon_{33}(E_2) - \varepsilon_{33}(E_1)}{E_2 - E_1}$ within an individual ferroelectric (FE) nanocrystal or single grain in a polycrystalline material. To demonstrate the efficacy of BCDI, we spatially resolved $d_{33}$ within the nanocrystal (Fig. 3) and identified the topological vortex core as a 1-dimensional filament spanning the volume of the FE nanocrystal (Fig. 3a, b). Cross-sectional views reveal $d_{33}$ enhancements near the core with values as large as ~1400 pC/N.

Our analysis shows that while the value of $d_{33}$ decays away from the vortex core towards experimentally reported bulk values, this decay is not uniform in all directions from the core (Fig. 3c). We observe two distinct trends in its spatial behavior away from the core along co-existing structural phase regions within the nanocrystal. Radial plots of $d_{33}$ near the core reveal an oscillatory behavior, indicating transitions across existing structural phase boundaries (Fig. 3d), where abrupt changes in the lattice strain due to polarization rotations are expected.

In order to understand the overall behavior of $d_{33}$ within the ferroelectric (FE) nanocrystal, we estimated a weighted average over 15 pixels (1 pixel = 6 nm) radially from the vortex core, which resulted in a power law variation (Fig. 4a). Power laws are observed in many natural phenomena and can arise from various mechanisms, including self-organization, critical phenomena, and multiplicative processes. Critical phenomena occur when a system is at the threshold of a transition between different states. In-operando BCDI shows high strains and large $d_{33}$ where structural phases coalesce (Fig. 4b).

The exponent α in the observed power law k r$^{-α}$ represents the scaling behavior of $d_{33}$ with distance from the vortex core (Fig. 4a). Specifically, it indicates how quickly $d_{33}$ decays with increasing distance from the core. Log-log plots give different values of the constant k and α in the range 1.75≤α≤ 3.75. For example, if α is small and the power law relationship holds over a wide range of distances, this suggests that $d_{33}$ decays relatively slowly with increasing distance from the vortex cores.

This could indicate that the presence of multiple vortices within the nanocrystal has a relatively uniform effect on the piezoelectric coefficient across the entire sample. On the other hand, if α is large and the power law relationship holds over a narrow range of distances, this suggests that $d_{33}$ decays rapidly with increasing distance from the vortex cores. In this case the presence of multiple vortices may have a more localized effect on the piezoelectric coefficient, and the behavior of $d_{33}$ may be strongly influenced by the geometry and distribution of the vortices within the nanocrystal.

Our MD simulations show that the applied electric field induces large changes in local strain (shown in Fig. 4c) in a nanodot with topological FE vortices, expected to lead to enhanced local piezoelectric response[19]. Figure 4d shows the distribution functions for the largest lattice parameter of the individual unit cells in the nanodot computed for different values of applied electric field. The field is applied perpendicular to one of the cube's faces. The calculations suggest that in the absence of an electric field the distribution function is very broad, which is attributed to the predominance of lower symmetry structural phases (labeled X in Fig. 1 and Fig 4d.) which cannot be resolved in the simulations (see also supplemental section-4). Electric field narrows the distribution function and shifts the maximum towards larger values. The equilibrium dipole patterns in the cross section of the nanodot originate from the delicate balance between the ferroelectric instability and the depolarization effect due to uncompensated surface charge[4,5].

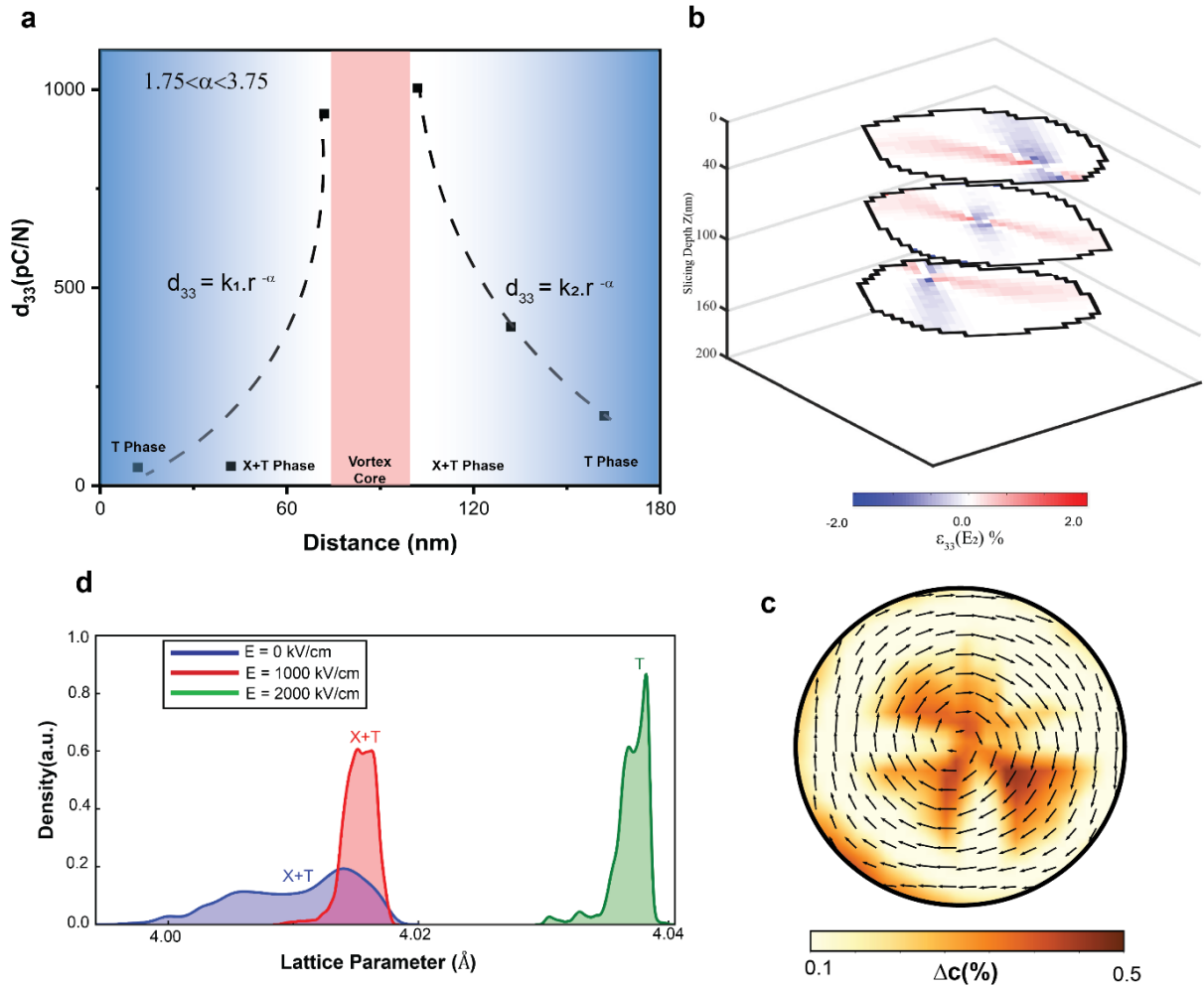

**Figure 4: a: Spatial and Directional Variations in Piezoelectric Response ($d_{33}$) and Strain ($\varepsilon_{33}$) in BTO Nanocrystal: Insights from Power Law Dependence and MD Simulations.** Power law depicting weighted averaged variation of $d_{33}$ with radial distance away from the vortex core. We power law dependence $d_{33} \sim k(r)^{-\alpha}$, with directional dependence of the exponent $\alpha$ and proportionality constant k. We notices variations of $\alpha$ from 1.75 and 3.75. The accuracy ($R^2$) of the fit varies from 0.85 and 0.95. **b:** 2D slices showing the spatial variation of the strain ($\varepsilon_{33}$) along the z-direction in BTO.. **c:** Cross section of the simulated nanodot. Color gives changes in the largest unit cell lattice parameter caused by the transformation of the FE vortex (shown by the black arrows) to a homogeneous polarization field. The transformation is induced by the application of 2000 kV/cm electric field. The largest $d_{33}$ occurs near the vortex core in agreement with BCDI data in the previous panel. **d:** Distribution functions for the largest of the three-unit cell lattice parameters computed at room temperature from MD under zero and finite electric fields.

The rotating pattern annihilates the depolarizing field but causes the unit cell dipoles to deviate from their tetragonal axes and giving origin to the low-symmetry phases revealed from the distribution function of (see also supplemental Fig. S6) Note that under zero field vortices exist in all three orthogonal cross sections of the nanodot (Fig. S5). Under applied field of 1000 kV/cm the vortex is preserved in the plane perpendicular to the direction of the applied field but is erased in the other planes (see Fig. S4). The field of 2000 kV/cm completely erases the vortex, thus eliminating the X phases, as observed previously[3]. Fig 4c gives the changes in the c-lattice parameter induced by the applied electric field of 2000 kV/cm in the cross section of a nanodot as obtained from MD simulations. The data reveal that the regions of enhanced strain change, or the piezoelectric response, are indeed the ones associated with the X-phase, which provides further support to the insight from experiment and is consistent with our phase field simulations that showed how the vortex disappears for large enough electric fields[3]. We note that the field in computations is likely to overestimate experimental field as we model a defect free nanodot with nearly fully unscreened surface charge. Thus, our computational data predict that an applied electric field induces large changes in local strain in nanostructures with topological FE vortices, expected to lead to enhanced local piezoelectric response[19].

We have demonstrated enhanced piezoelectric response in the vicinity of FE topological vortex core. Interestingly, we observe extremely small values of the FE spontaneous polarization ($P_s$) near the core. This further suggest that the correlation between $P_{max}$ and $d_{33}$ established in Fig.1a for bulk BT compounds is different for undoped BTO at the nanoscale. For data points extracted as a walk away from the vortex core (Fig. 1b) in the nanocrystal, we observe that the largest $d_{33}$ values are associated with the smallest $P_s$, unlike bulk observations.

The above trends suggests that the vortex core can be considered as a nanoscopic structural analogue to the compositional triple point in macroscopic phase diagrams. In this analogue, the vortex core is a relatively "soft" region acting as a parent structural phase where the tetragonal and X phases intersect. Under an electric field, these "soft" vortex core regions will produce extremely large strain responses in FE materials under relatively small electric fields. We can only speculate on the validity of $d_{33}$ ~1/P for both bulk and nanoscale. In contrast to equilibrium, the nanoscale structures are metastable and one could think in terms of a microstructural phase diagram with the vortex core serving the role of a triple point where the transition is essentially continuous. This raises the question of whether we can postulate the existence of tricritical points in this microstructural phase diagram where the nature of the transition changes away from the core. Given that our observations are at the level of microstructure, an interesting question is how the contributions from a density of vortices integrate to give rise to a $d_{33}$ up to 2100 pC/N observed in nanoparticles of BTO ? We consider only an isolated vortex and hence challenging issues arise when an assembly of interacting vortices are studied. In future, upgrades to synchrotrons and X-ray Free Electron Lasers will enable the investigation of material structures at extremely high resolutions, approaching atomic dimensions. This progress will enable high-resolution imaging of materials in operando conditions in 3D. These advances will be possible with the development of BCDI techniques, which can provide researchers with insights into the behavior of ferroelectrics and allowing to engineer microstructures that harness large $d_{33}$ at the nanoscale.

In summary, our results suggest a paradigm for nano-electromechanical device engineering that utilizes large piezoelectric response in nanoscale ferroelectric systems. As FE vortices are inherent to nanoscale ferroelectrics[3–5], we expect our findings to also apply to other classes of low dimensional ferroelectrics.

**Materials and Methods**

**External electric field control:** We used the commercial Precision-II ferroelectric tester (Radiant technology) to carry out programmable application of the electric field and *in-situ* feedback monitoring of the FE sample. The system is composed of elements such as low noise and high sensitivity electro-meter amplifiers, precision potentiometer, circuit integrated reference capacitors, and integrating capacitors, and leakage impedance compensators. We performed ferroelectric tests in compliance with the IEEE standard 180 and simultaneously drove the sample with stable pulses through an integrated function generator. The system can monitor global hysteresis, leakage, induced charge, resistivity and other properties during the experiment with maximum charge and voltage resolution of 0.8 fC and 76 μV respectively at capture rate of 0.5 μs. The maximum frequency of applied pumping pulses with continuous hysteresis measurements feedback is 250 kHz. The testing system is fully integrated and programmable, thus, enabling algorithmic on-line control of the experimental parameters based on the readings.

**Ferroelectric capacitor:** The detailed diagram of the ferroelectric capacitor used in our measurements is illustrated in Supplementary Fig. S3. The BTO nanocrystal was first pre-characterized with laboratory X-ray diffraction (XRD). The sample shows expected tP5 crystal structure with high quality of crystallinity as seen in the XRD result in Supplementary Fig. S1. The BTO nanocrystals were thoroughly mixed with a conductive polymer matrix prior to the BCDI measurements. The volume fraction of incorporated BTO nanocrystals was 15%. The polymer matrix was prepared in advance by mixing liquid bisphenol A based epoxy with conducting carbon nanocrystals (40% by volume). The composite solution was transferred into a template with dimensions $2 \times 2 \times 1$ (mm ×mm× mm) and cured with a commercially available agent in an oven first at 90 for 1 h and then at 60° for 4 h. Gold electrodes were sputtered on corners of two sides of the specimen and thin conductive Kapton film was attached with insulating surfaces facing to the outward direction. Dispersed particles are expected to form conducting chains in epoxy polymers. This above methodology was performed to produce the current directly to embedded BTO nanocrystal. During the experiments voltage scans of 30 cycles (Supplementary Fig. S3) were performed with monitoring of evolution of diffraction intensities to differentiate the active (linked in the conductive chain) and inactive (insulated in the epoxy matrix) BTO nanocrystals. When the BTO nanocrystal has become isolated from the conducting particles, the measured coherent diffraction intensities show no corresponding variations with the applied voltage.

**Experiment and phase retrieval procedure:** We used a Si (111) monochromator sector 34-ID-C of the Advanced Photon Source to select coherent X-ray photons with an energy of 9.0 keV with a 1~eV bandwidth monochromaticity and transverse coherence length of 0.7 μm. We used Kirkpatrick—Baez (KB) mirrors to focus the X-rays onto the sample down to a beam size of 700 nm by 700 nm. A schematic showing the principle of Bragg coherent diffraction experiment is depicted in Fig.1d. We used a motorized arm to position a Medipix2 CMOS X-ray detector around the diffraction sphere. We align the detector position to the outgoing (002) Bragg reflection from the BTO. We placed the detector at a distance of 1.2 m from the sample to zoom into the interference fringes in the diffraction pattern. An evacuated flight tube placed in the sample-to-detector path helps to decrease the loss of photons scattered in the air. The use of an evacuated flight tube, high sensor gain, as well as photon counting mode of the detector, are crucial for such photon-starving techniques as nanoscale Bragg coherent diffractive imaging. The rocking curve was collected as a series of 2D diffraction patterns in the vicinity of the BTO {002} Bragg peak with the scanning range of $\Delta\theta = \pm0.25°$ about the Bragg peak origin. Throughout a single rocking curve, a total scan of 150 patterns was collected. The dataset for the virgin state $E_1$ was collected before cycling the functional capacitor. The following state under applied electric field $E_2$ was recorded after 30 cycles of applied and released electric field. This allowed us to make sure that the system underwent a phase-transition under an applied electric field.

**Reconstruction procedure:** We invert the recorded 3D diffraction patterns in reciprocal space into a direct space 3D image of the nanocrystal after the phase $\phi(r)$ and amplitude $\rho_0(r)$ of the complex wave-field $\rho(r) = \rho_0(r)e^{i\phi(r)}$ are iteratively retrieved. The reconstructed phases and displacement field $u_{002}(r)$, for a given reciprocal lattice vector $G_{002}$ obey the following relationship $\phi(r) = G_{002} \cdot u_{002}(r)$. We estimate the shape and size of the nanocrystal from the isosurface of reconstructed Bragg electron density $\rho(r)$ (see Fig 2).

For a given external electric field, real-space images of the BTO nanocrystals were reconstructed with approximately 15 nm spatial resolution as determined by the phase retrieval transfer function (PRTF)[28](see Fig.S9). The reconstruction allows us to conveniently slice through the volume of the nanoparticle at a given external electric field and analyze the topological FE structure (Fig.2) of BTO and signatures the piezoelectric response (Fig.4). Iterative phase retrieval algorithms based on Fienup's Hybrid Input-Output (HIO) method with additional randomized over-relaxation ($O_R$) were utilized. Inverting the diffraction data is a critical step that uses a computer algorithm that takes advantage of internal redundancies when the measurement points are spaced close enough together to meet the oversampling requirement. The first step is to postulate a 3D support volume in which all the sample density will be constrained to exist. These methods impose a backward and forward Fourier transform between the reciprocal space and the real-space, with the support constraint imposed in the latter and an intensity mask constraint in the former. More on the reconstruction procedure is detailed in the supplemental.

**Theoretical Computations**

**MD Simulations:**

For our first-principles Hamiltonian[21] the degrees of freedom include local modes, $u_i$, that are proportional to the dipole moment in the unit cell, and strain variable tensors, $\eta_i$ (in Voigt notations), that are responsible for mechanical deformations of a unit cell. The total energy of the effective Hamiltonian is given by

$$H^{Tot} = H^{FE}(u_i) + H^{elas}(\eta_i) + H^{FE-elas}(u_i,\eta_i) + H^{elec}(u_i) \qquad \text{Eq.(2)}$$

Here, $H^{FE}$ is the energy associated with the ferroelectric local modes and includes contributions from the dipole-dipole interactions, short-range interactions, and on-site self-energy as defined in Ref.[29]. The second term, $H^{elas}$ is the elastic energy associated with the unit cell deformations. $H^{FE-elas}$ is the energy contribution due to the interactions between the ferroelectric local modes and the strain. The last term in Eq. (2) $H^{elec} = Z_i^* \Sigma_i E. u_i$ is the interaction energy between the local modes and an external electric field, E, while Z* is the local mode Born effective charge. This Hamiltonian reproduces correctly the complex sequence of phase transitions in Ba(Zr,Ti)O$_3$ solid solution in wide compositional range[21]. For MD simulations the Newton equation of motion are integrated numerically using 1 fs time step. The NVT ensemble is simulated using Evans-Hoover thermostat[30]. For a given temperature the simulations are run for 100 000 MD steps, the last half of which are used to compute thermodynamical average values. Such an approach has previously been used to simulate dynamical and equilibrium properties of ferroelectrics and their nanostructures[17,18,20,22]. The dc electric field was applied/removed by sequential increase/decrease from 0 to the 2000 kV/cm.

## Author Information

Reprints and permissions information is available at www.nature.com/reprints. The authors declare no competing financial interests. Readers are welcome to comment on the online version of the paper. Correspondence and requests for materials should be addressed to E.F. (fohtue@rpi.edu).

## Acknowledgments

This work is supported by the US Department of Defense, Airforce Office of Scientific Research (AFOSR) under award No. FA9550-23-1-0325 (Program Manager: Dr. Ali Sayir) and the US Department of Energy, (DOE) Office of Science under grant No. DE-SC0023148. This research used resources of the Advanced Photon Source (APS), a U.S. Department of Energy (DOE) Office of Science User Facility operated for the DOE Office of Science by Argonne National Laboratory (ANL) under contract No. DE-AC02-06CH11357. R.K, F.S.T, and I.P. acknowledge support from the U.S. Department of Energy, Office of Basic Energy Sciences, Division of Materials Sciences and Engineering under Grant No. DE-SC0005245.

## Conflicts of interest

The authors declare no conflict of interest.

## Author contributions

D.K, LT and E.F. designed and performed the experiments. N.N, X.S, D.K., and E.F analyzed the data. R.K, F.S.T, BK and I.P. performed MD simulations, and first-principles calculations. All authors contributed to the writing and editing of the manuscript.

**Supplementary Materials**

Supplementary Text

Figs. S1 to S14

Table S1